%% file: ms.tex
\documentclass[conference, final]{IEEEtran}
\usepackage{graphicx}
\graphicspath{{./}}
\usepackage{subfig}
\usepackage{multirow}
\usepackage{type1cm}
\usepackage{amsmath}
\usepackage{booktabs, threeparttable}
\usepackage{color}
\usepackage{enumitem}
\usepackage{url}
\usepackage{listings}
\usepackage{caption}
\usepackage{algpseudocode}
\usepackage{tikz}
\usepackage{pgfplots}
\usepackage{float}
\usetikzlibrary{matrix}
\usepackage[ruled,linesnumbered,lined,boxed,commentsnumbered]{algorithm2e}

\usepackage{epigraph}

\usepackage{etoolbox}
\makeatletter
\patchcmd{\epigraph}{\@epitext{#1}}{\itshape\@epitext{#1}}{}{}
\makeatother

\definecolor{mygreen}{rgb}{0,0.6,0}
\definecolor{mygray}{rgb}{0.5,0.5,0.5}
\definecolor{mymauve}{rgb}{0.58,0,0.82}

\newcounter{lemma_counter}
\newcounter{theory_counter}
\begin{document}
\title{\bf{Concurrent CPU-GPU Task Programming using Modern C++}}

\author{\IEEEauthorblockN{anonymous\IEEEauthorrefmark{1}}
\IEEEauthorblockA{
\IEEEauthorrefmark{1}affiliation left empty for blind review}
}

\author{\IEEEauthorblockN{Tsung-Wei Huang\IEEEauthorrefmark{1} and
Yibo Lin\IEEEauthorrefmark{2}}
\IEEEauthorblockA{\IEEEauthorrefmark{1}Department of Electrical and Computer Engineering, University of Utah}
\IEEEauthorblockA{\IEEEauthorrefmark{2}Department of Computer Science, Peking University}}

\date
\small \maketitle 


\begin{abstract}

In this paper, we introduce Heteroflow,
a new C++ library to help developers quickly
write parallel CPU-GPU programs using task dependency graphs.
Heteroflow leverages the power of modern C++ and 
task-based approaches to enable efficient implementations 
of heterogeneous decomposition strategies.
Our new CPU-GPU programming model allows users to express 
a problem in a way that adapts
to effective separation of concerns and expertise encapsulation.
Compared with existing libraries, 
Heteroflow is more cost-efficient in performance scaling,
programming productivity, and solution generality.
We have evaluated Heteroflow on two real applications
in VLSI design automation and demonstrated the performance scalability
across different CPU-GPU numbers and problem sizes.
At a particular example of VLSI timing analysis with million-scale tasking,
Heteroflow achieved $7.7\times$ runtime speed-up (99 vs 13 minutes) 
over a baseline
on a machine of 40 CPU cores and 4 GPUs.

\end{abstract}


\section{Introduction}
\label{sec::introduction}

Modern parallel applications in machine learning, data analytics, 
and scientific computing typically
consist of a heterogeneous use of both 
central processing units (CPUs) and 
graphics processing units (GPUs)~\cite{Vetter_18_01}.
Writing a parallel CPU-GPU program 
is never an easy job,
since CPUs and GPUs have fundamentally different architectures
and programming logic.
To address this challenge,
the parallel computing community has investigated
many programming libraries
to assist developers with quick access to massively parallel
and heterogeneous computing resources
using minimal programming effort~\cite{CUDA, OpenCL, hiCUDA, Ompss, OpenMPC, OpenACC, StarPU, SYCL, HPX, PaRSEC}.
In particular, 
hybrid multi-CPU multi-GPU systems are driving high demand
for new heterogeneous programming techniques 
in support for more efficient CPU-GPU collaborative computing~\cite{Mittal_15_01}.
However, related research remains nascent,
especially on the front of leveraging modern C++ to 
achieve new programming productivity and performance scalability
that were previously out of reach~\cite{Huang_19_01}.

\begin{figure}[h]
  \centering
  \centerline{\includegraphics[width=1.\columnwidth]{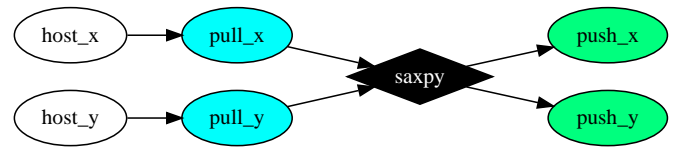}}
  \caption{A saxpy (``single-precision A·X plus Y") task graph using two \textit{host} tasks to create two data vectors, two \textit{pull} tasks to send data to a GPU, a \textit{kernel} task to offload the saxpy computation to the GPU, and two \textit{push} tasks to push data from the GPU to the host.}
  \label{fig::saxpy}
\end{figure}

The Heteroflow project addresses a long-standing question:
``\textit{how can we make it easier for C++ developers to 
write efficient CPU-GPU parallel programs?}"
For many C++ developers,
achieving high performance on a hybrid CPU-GPU system can
be tedious.
Programmers have to overcome complexities arising out of
concurrency controls, kernel offloading, scheduling,
and load-balancing
before diving into the real implementation of a heterogeneous
decomposition algorithm.
Heteroflow adopts a new \textit{task-based} programming model using modern C++
to address this challenge.
Consider the canonical saxpy (A·X plus Y) example in Figure \ref{fig::saxpy}.
Each Heteroflow task belongs to one of 
\textit{host}, \textit{pull}, \textit{push}, and \textit{kernel} tasks;
a host task runs a callable object on any CPU core (``the host"),
a pull task copies data from the host to a GPU (``the device"),
a push task copies data from a GPU to the host, and
a kernel task offloads computation to a GPU.
Figure \ref{fig::saxpy} explains
the saxpy task graph in Heteroflow's graph language.

\begin{lstlisting}[language=C++,label=saxpy_code,caption={\small Heteroflow code of Figure \ref{fig::saxpy}.}]
__global__ void saxpy(int n, int a, int *x, int *y){
  int i = blockIdx.x * blockDim.x + threadIdx.x;
  if (i < n) y[i] = a*x[i] + y[i];
}

const int N = 65536;
vector<int> x, y;

hf::Executor executor;
hf::Heteroflow G;

auto host_x = G.host([&](){ x.resize(N, 1); });
auto host_y = G.host([&](){ y.resize(N, 2); });
auto pull_x = G.pull(x);
auto pull_y = G.pull(y);
auto kernel = G.kernel(saxpy, N, 2, pull_x, pull_y)
               .block_x(256)
               .grid_x((N+255)/256)
auto push_x = G.push(pull_x, x);
auto push_y = G.push(pull_y, y);

host_x.precede(pull_x); 
host_y.precede(pull_y); 
kernel.precede(push_x, push_y)
      .succeed(pull_x, pull_y);

auto future = executor.run(hf);
\end{lstlisting}

Listing \ref{saxpy_code} shows the Heteroflow code
that implements the saxpy task graph in Figure \ref{fig::saxpy}.
The code \textit{explains itself}.
The program creates a task dependency graph of 
two host tasks, two pull tasks, one kernel task, and two push tasks.
The kernel task binds to a saxpy kernel written in CUDA~\cite{CUDA}.
The dependency links form constraints that conform to Figure \ref{fig::saxpy}.
Heteroflow provides an \textit{executor} interface to perform automatic 
parallelization of a task graph scalable to manycore CPUs and GPUs.
There is no explicit thread managements or fine-grained concurrency controls
in the code.
Our design principle is to let users write
\textit{simple}, \textit{expressive}, and \textit{transparent} parallel code.
Heteroflow explores a minimum set of core routines 
that are sufficient enough for users to implement
a broad set of heterogeneous computing algorithms.
Our task application programming interface (API) is not only flexible 
on the user front, 
but also extensible with the future evolution of C++ standards
and heterogeneous architectures.
We summarize our contributions as follows:

\begin{itemize}[leftmargin=*]\itemsep=2pt

\item \textbf{Programming model}. 
We develop a new parallel CPU-GPU programming model
to assist developers with efficient access to
heterogeneous computing resources.
Our programming model allows users to express a problem
with effective separation of concerns
and expertise encapsulation.
Developers can work at a suitable level of granularity 
for writing scalable applications that is commensurate with 
their domain knowledge.

\item \textbf{Transparency}.
Heteroflow is transparent.
Developers need not to deal with standard concurrency mechanisms
such as threads and fine-grained concurrency controls, 
that are often tedious and hard to program correctly.
Instead, our system runtime abstracts these problems from developers 
and tackles many of the hardest parallel and heterogeneous computing
details, notably resource allocation, CPU-GPU co-scheduling,
kernel offloading, etc.

\item \textbf{Expressiveness}.
We leverage modern C++ to design an expressive API that empowers
users with explicit graph construction and refinement to fully exploit
task parallelism in their applications.
The expressive power also lets developers perform rather
a lot of work without writing a lot of code.
Our user experiences lead us to believe that,
although it requires some effort to learn,
most C++ programmers can master our APIs and 
apply Heteroflow to their jobs in just a few hours.

\end{itemize}

We have applied Heteroflow to two real applications,
timing analysis and cell placement, in large-scale circuit design automation
and demonstrated the performance scalability across 
different numbers of CPUs, GPUs, and problem sizes.
We believe Heteroflow stands out as a unique tasking library considering
the ensemble of software tradeoffs and architecture decisions we have made.
With that being said,
different programming libraries and frameworks have their pros and cons,
and deserve a particular reason to exist.
Heteroflow aims for a higher-level alternative 
in modern C++ domain.


\section{Motivation}

\begin{figure}[h]
  \centering
  \centerline{\includegraphics[width=1.\columnwidth]{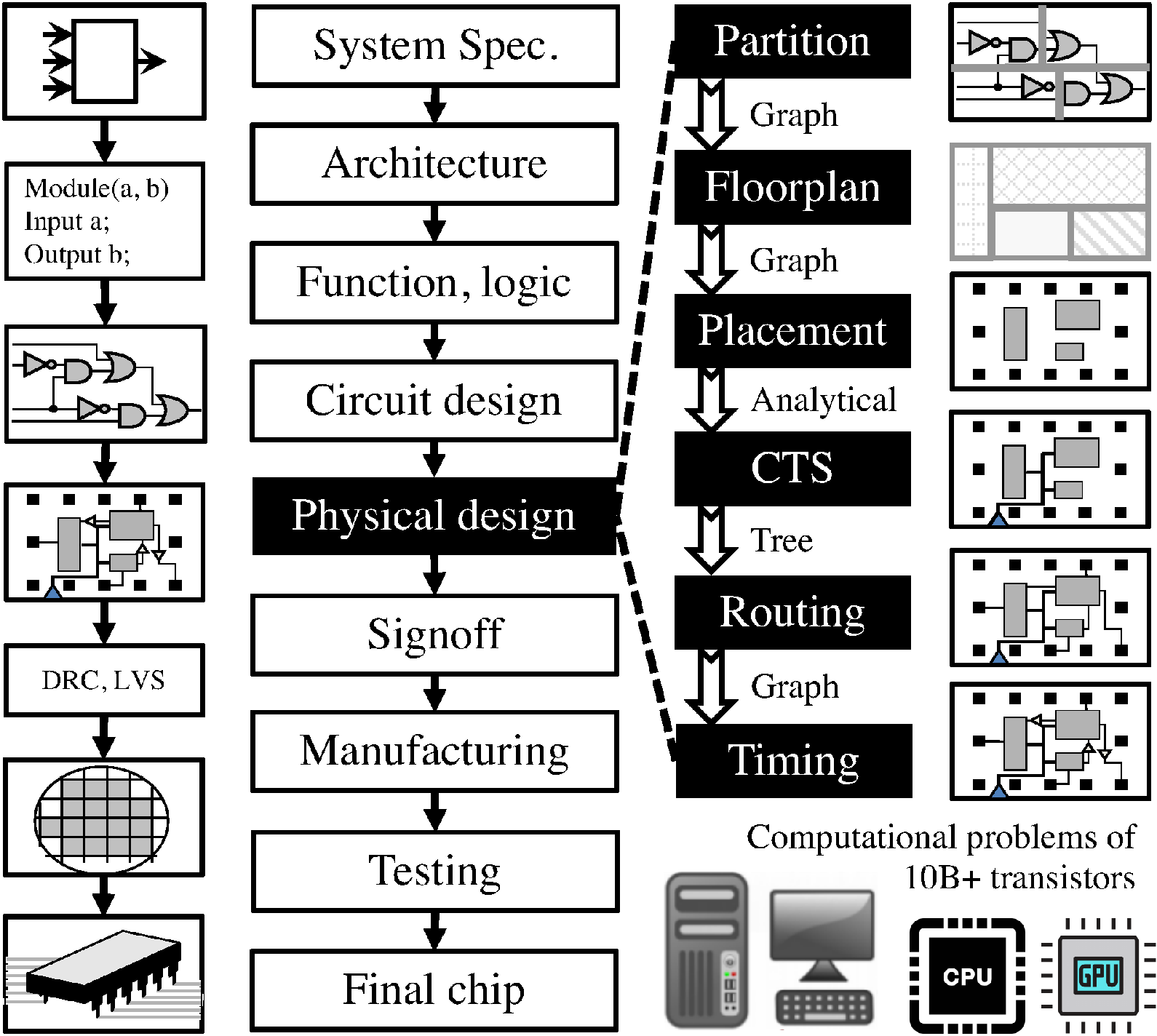}}
  \caption{A typical VLSI design automation flow with a highlight on the physical design stage. Heteroflow is motivated to address the ever-increasing computational need of modern CAD tools.}
  \label{fig::cad}
\end{figure}

Heteroflow is motivated by our research projects to develop
efficient computer-aided design (CAD) tools for
very large scale integration (VLSI) design automation.
CAD has been an immensely successful field in
assisting designers in implementing VLSI circuits with billions
of transistors.
It was on the forefront of computing around 1980 and 
has fostered many prominent problems and algorithms in computer science.
Figure \ref{fig::cad} demonstrates a conventional VLSI CAD flow
with a highlight on physical design.
Due to the ever-increasing design complexity,
the recent CAD community is driving the need for 
hybrid CPU-GPU computing to keep tool performance up with 
the technology scaling~\cite{Stok_14_01, Lu_18_01}.

\subsection{Challenge 1: Vast and Complex Dependencies}

Computational problems in CAD are extremely complex and have many
challenges that normal software developments do not have.
The biggest challenge to develop parallel CAD tools is the vast
and complex task dependencies.
Before evaluating an algorithm, 
a number of logical and physical information
must arrive first. 
These quantities are often dependent to each other and 
are expensive to compute.
The resulting task dependency graph in terms of 
encapsulated function calls
can be very large.
For example, a million-gate design can produce a graph
of billions of tasks and dependencies 
that takes several days to accomplish~\cite{Huang_19_01}.
However, such difficulty does not prevent CAD tools from parallelization, 
but highlights the need of new tasking frameworks to implement
efficient parallel decomposition strategies especially
with CPU-GPU collaborative computing~\cite{Lu_18_01}.

\subsection{Challenge 2: Extensive Domain Knowledge}

Developing a parallel CAD algorithm requires
deep and broad domain knowledge across circuits, modeling,
and programming to fully exploit parallelism.
The compute pattern is highly irregular and unbalanced,
requiring very strategic collaboration between CPU and GPU.
Developers often need direct access to native GPU programming libraries 
such as CUDA and OpenCL to handcraft the kernels with 
problem-specific knowledge~\cite{CUDA, OpenCL}.
Existing frameworks that provide high-level abstraction 
over kernel programming always come with restricted applicability,
preventing CAD engineers from using many new powerful features of
the native libraries.
Our domain experience concludes that
despite nontrivial GPU kernels,  
\textit{what makes concurrent CPU-GPU programming an enormous challenge 
is the vast and complex surrounding tasks, most notably
the resource controls on multi-GPU cards, CPU-GPU co-scheduling, tasking,
and synchronization.}

\subsection{Need for a New CPU-GPU Programming Solution}

Unfortunately, most parallel CPU-GPU programming solutions 
in CAD tools are hard-coded~\cite{Stok_14_01, Lu_18_01}.
Developers are ``heroic programmers" to handcraft every detail
of a heterogeneous decomposition algorithm
and explicitly decide which part
of the application runs on which CPU and GPU.
While the performance is acceptable,
it is too expensive to maintain the codebase and 
scale to new hardware architectures.
Some recent solutions adopted directive-driven models 
such as OpenMP GPU and OpenACC 
particularly for data-intensive algorithms~\cite{OpenMPC, OpenACC}.
However, these approaches cannot handle dynamic workloads
since compilers have limited knowledge to 
annotate runtime task parallelism and dynamic dependencies.
In fact, frameworks at functional level are more favorable 
due to the flexibility in runtime controls and on-demand tasking.
Nevertheless,
most libraries on this front are disadvantageous
from an ease-of-programming standpoint~\cite{Mittal_15_01}.
Users often need to sort out many distinct notations and library details
before implementing a heterogeneous algorithm~\cite{Beri_16_01}.
Also, a lack of support for modern C++ largely inhibits 
the programming productivity and performance scalability~\cite{Huang_19_01, Huang_22_01}. 
After many years of research, 
we and our industry partners conclude
the biggest hurdle to program the power of collaborative 
CPU-GPU computing is a suitable \textit{task programming library}.
Whichever model is used, 
understanding the structure of an application is critical. 
Developers must explicitly consider possible data or task parallelism 
of their problems and leverage domain-specific knowledge
to design effective decomposition strategies for parallelization.
At the same time,
the library runtime removes the burden
of low-level jobs from developers to improve programming productivity
and transparent scalability.
To this end, our goal is to address these challenges
and develop a general-purpose tasking interface for concurrent
CPU-GPU programming.

\section{Heteroflow}

In this section, we discuss the programming model
and runtime of Heteroflow.
We will cover important technical details 
that support the software architecture of Heteroflow.

\epigraph{Heteroflow aims to help C++ developers quickly write 
CPU-GPU parallel programs and implement 
efficient heterogeneous decomposition strategies using task-based models.}{--- \textup{Heteroflow's Project Mantra}}

\subsection{Create a Task Dependency Graph}

Heteroflow is \textit{object-oriented}.
Users can create multiple task dependency graph objects
each representing a unique parallel decomposition in an application.
A task dependency graph is a \textit{directed acyclic graph} (DAG)
with nodes and edges representing tasks and dependency constraints, respectively.
Each task belongs to one of the four categories:
\textit{host}, \textit{pull}, \textit{push}, 
and \textit{kernel}.

\subsubsection{Host Task}

A host task is associated with a \textit{callable} object
which can be a function object, binding expression, functor, or a lambda expression.
The callable is invoked at runtime by a CPU thread to run on a CPU core.
Listing \ref{HostTaskCode} gives an example of creating a host task.
In most applications, the callable is described in C++ lambda
to construct a \textit{closure} inline in the source code.
This property allows host task to enable efficient lazy evaluation
and capture any data whether it is declared in a local block or 
flat in a global scope, largely facilitating the ease of programming.

\begin{lstlisting}[language=C++,label=HostTaskCode,caption={Creates a host task.}]
hf::Heteroflow hf;
auto host = hf.host(
  [] () { cout << "task runs on a CPU core"; }
);
\end{lstlisting}

Each time users create a task,
the heteroflow object adds a node to its task graph
and returns a \textit{task handle} to users.
A task handle is a lightweight class object
that wraps a pointer to a graph node.
The purpose of this extra layer
is to provide an extensible mechanism for users to
modify the task attributes
and, most importantly, prevents users from 
direct access to the internal graph storage
which can easily introduce undefined behaviors.
Each node has a general-purpose polymorphic function wrapper 
to store and invoke different callables according to a task type.
A task handle can be empty, often used as a \textit{placeholder}
when it is not associated with a graph node.
This is particularly useful when a task content
cannot be decided until a certain point during the program execution,
while the task storage needs preallocation at programming time.
These properties are applicable to all task types.

\subsubsection{Pull Task}

A pull task lets users \textit{pull} data from the host to the device.
The exact GPU to perform this memory operation is decided by the scheduler
at runtime.
Developers should think separately which part of their applications 
runs on which space,
and decompose them with explicit task construction.
Since most GPU memory operations are expensive compared to CPU counterparts,
Heteroflow splits the execution of a GPU workload into three operations,
host-do-device (H2D) input transfers, launch of a kernel, and 
device-to-host (D2H) output transfers, 
to enable more task overlaps.
Pull task adopts this strategy to help 
users manage the tedious details in H2D data transfers.
At the same time, it presents an effective abstraction
of which the scheduler can take advantage to perform various optimizations
such as automatic GPU mapping, streaming, and memory pooling.

\begin{lstlisting}[language=C++,label=PullTaskCode,caption={Creates two pull tasks.}]
vector<int> data1(100);
float* data2 = new float[10];
auto pull1 = hf.pull(data1);
auto pull2 = hf.pull(data2, 10);
\end{lstlisting}

Listing \ref{PullTaskCode} gives an example of creating two pull tasks
to transfer data from the host to the device.
The first pull task operates on a C++ vector of integer numbers and
the second pull task operates on a raw data block of real numbers.
Heteroflow employs the C++20 \textit{span} syntax to 
implement the pull interface.
The arguments forwarded to the pull method must conform to the constructor
of \texttt{std::span}.
In fact, we have investigated many possible data representations
and decided to use span because of its lightweight abstraction
for describing a contiguous sequence of objects.
A span can easily convert to a C-style raw data view 
that is acceptable by 
most GPU programming libraries~\cite{CUDA, OpenCL, OpenGL}.
Sticking with C++ standard also keeps the core of Heteroflow portable and
minimizes the rate of change required for our data interface.

\begin{lstlisting}[language=C++,numbers=left,xleftmargin=1.8em,label=PullTaskImplCode,caption={Implementation details of the pull task.}]
template <typename... ArgsT>
auto PullTask::pull(ArgsT&&... args) {
  get_node_handle().work = [
    t=StatefulTuple(forward<ArgsT>(args)...)
  ] (Allocator& a, cudaStream_t s) mutable {
    auto h_span = make_span_from_tuple(t);
    auto h_data = h_span.data();
    auto h_size = h_span.size_bytes();
    auto d_data = a.allocate(h_size);
    cudaMemcpyAsync(
      d_data, h_data, h_size, H2D, s
    );
  };
  return *this;
}
\end{lstlisting}

Listing \ref{PullTaskImplCode} highlights the core implementation 
of the pull task based on CUDA.
\footnote{While the current implementation is based on CUDA,
our task interface can accept other 
GPU programming libraries~\cite{OpenCL}.}
To be concise, 
we omit details such as error checking and auxiliary functions.
The pull task forms a closure that captures the arguments 
in a custom tuple by which we enable \textit{stateful} task execution
(line 4).
For instance, in Listing \ref{saxpy_code},
the change made by the host task \texttt{host\_x} on the data vectors 
must be visible to the pull task \texttt{pull\_x}.
The stateful tuple wraps references in objects to
keep state transition consistent between dependent tasks.
Maintaining a stateful transition is a backbone of Heteroflow.
Developers can carry out fine-grained concurrency 
through decomposition and enforce dependency constraints
to keep the logical relationship between task data.
In terms of arguments,
the runtime passes a memory allocator and a CUDA stream
to the closure (line 5).
The allocator is a pooled resource for reducing
GPU memory allocation overhead
and
the CUDA stream is a sequenced mechanism for interleaving GPU operations~\cite{CUDA}.
A key motivation behind this design is 
to support multi-GPU computing.
Both the memory allocator and stream are specific to a GPU context
which is decided by the scheduler at runtime.
Finally,
we create a span from the stateful tuple and 
enqueue the data transfer operation to the stream (line 6:12).

\subsubsection{Push Task}

A push task lets users \textit{push} data associated with a pull task
from the device to the host.
The code snippet in Listing \ref{PushTaskCode} creates
two push tasks that operate on the pull tasks 
in Listing \ref{PullTaskCode}.
The arguments consist of two part, 
a source pull task of device data and
the rest to construct a \texttt{std::span} object for the target.
Similar to Listing \ref{PullTaskCode},
the first push task operates on an integer vector and
the second push task operates on a raw data block of floating numbers.
Push task is stateful.
Any runtime change on the arguments that were used to construct
a pull task will reflect on its execution context.
This property allows users to create stateful Heteroflow graphs
for efficient data management between concurrent CPU and GPU tasks.

\begin{lstlisting}[language=C++,label=PushTaskCode,caption={Creates two push tasks from the two pull tasks in Listing \ref{PullTaskCode}.}]
auto push1 = hf.push(pull1, data1);
auto push2 = hf.push(pull2, data2, 10); 
\end{lstlisting}

\begin{lstlisting}[language=C++,numbers=left,xleftmargin=1.8em,label=PushTaskImplCode,caption={Implementation details of the push task.}]
template <typename... ArgsT>
auto PushTask::push(PullTask p, ArgsT&&... args){
  get_node_handle().work = [
    src=p,
    t=StatefulTuple(forward<ArgsT>(args)...)
  ] (cudaStream_t s) mutable {
    auto h_span = make_span_from_tuple(t);
    auto h_data = h_span.data();
    auto h_size = h_span.size_bytes();
    auto d_data = src.device_data();
    cudaMemcpyAsync(
      h_data, d_data, h_size, D2H, s
    );
  };
  return *this;
}
\end{lstlisting}

Listing \ref{PushTaskImplCode} highlights the core implementation
of the push task based on CUDA.
The push task captures the argument list in the same way 
as the pull task to form a stateful closure (line 5).
The execution context
creates a span from the target and extracts the device data
from the source pull task (line 7:10).
Finally,
we enqueue the data transfer operation to a CUDA stream 
passed by the scheduler at runtime (line 11:13).
This CUDA stream is guaranteed to live in the same GPU context as
the source pull task.
In short, Heteroflow uses pull tasks and push tasks
to perform H2D and D2H data transfers.
Users explicitly specify the data to transfer between
CPU and GPU, 
and encode these tasks in a graph to exploit task parallelism.
They never worry about the underlying details of resource allocation
and GPU placement.

\subsubsection{Kernel Task}

A kernel task offloads computation from the host to the device.
Heteroflow empowers users with explicit kernel programming
using native CUDA toolkits.
We never try hard to develop another C++ kernel programming framework
that often comes with restricted applicability
and performance portability.
Instead, 
users leverage their domain knowledge with the highest degree of freedom
to implement their kernel algorithms,
while leaving task parallelism to Heteroflow.
Listing \ref{KernelTaskCode} gives an example of
creating two kernel tasks that offload two given CUDA kernel functions
to the device using the pull tasks created in Listing \ref{PullTaskCode}.
The first kernel task operates on \texttt{kernel1} 
with data from \texttt{pull1}.
The second kernel task operates on \texttt{kernel2}
with data from \texttt{pull2}.
Both tasks configure 256 CUDA threads in a block.
Kernel functions are not obligated to take any Heteroflow-specific objects.
This largely increases the portability and testability
of Heteroflow,
especially for applications that heavily use third-party kernel functions
written by domain experts.

\begin{lstlisting}[language=C++,label=KernelTaskCode,caption={Creates two kernel tasks that operate on the two pull tasks in Listing \ref{PullTaskCode}.}]
__global__ void kernel1(int* data, int N);
__global__ void kernel2(float* data, int N);
auto k1 = hf.kernel(kernel1, pull1, 100)
            .grid_x(N/256)
            .block_x(256);
auto k2 = hf.kernel(kernel2, pull2, 10);
            .grid(N/256, 1, 1)
            .block(256, 1, 1);
\end{lstlisting}

\begin{lstlisting}[language=C++,numbers=left,xleftmargin=1.8em,label=KernelTaskImplCode,caption={Implementation details of the kernel task.}]
template <typename F, typename... ArgsT>
auto KernelTask::kernel(F&& f, ArgsT&&... args) {
  gather_sources(args...);
  get_node_handle().work = [
    k=*this, f=forward<F>(f),
    t=StatefulTuple(forward<ArgsT>(args)...)
  ] (cudaStream_t s) mutable {
    k.apply_kernel(s, f, t);
  };
  return *this;
}

template <typename T>
auto KernelTask::gather_sources(T&&... tasks) {
  if constexpr(is_pull_task<T>) {
    (get_node_handle().add_sources(tasks), ...);
  }
}

template<typename F, typename T>
auto KernelTask::apply_kernel(
  cudaStream_t s, F f, T t
) {
  const auto N = tuple_size<T>::value;
  apply_kernel(s, f, t, make_index_sequence<N>{});
}

template<typename F, typename T, size_t ... I>
auto KernelTask::apply_kernel(
  cudaStream_t s, F f, T t, index_sequence<I ...>
) {
  auto& h = get_node_handle();
  f<<<h.grid, h.block, h.shm, s>>>(
    convert(get<I>(t))...
  );
}
\end{lstlisting}

Listing \ref{KernelTaskImplCode} highlights the core implementation
of the kernel task.
The kernel method takes a kernel function written in CUDA 
and the rest arguments to invoke the kernel
(line 1:2).
The arity must match in both sides.
A key difference between Heteroflow and existing models
is the way we establish data connection --
\textit{we use pull tasks as the gateway rather than raw pointers}.
This abstraction largely improves safety and transparency
in scaling graph execution to multiple GPUs.
From the input argument list,
we gather all relevant pull tasks to this kernel
(line 3 and line 13:18)
and let the scheduler perform automatic device placement.
Similar to push and pull tasks,
we capture the argument list in a stateful tuple (line 6)
and use two auxiliary functions to invoke the kernel 
from the tuple (line 20:36).
All the runtime changes on the arguments 
will reflect on the execution context of the kernel.

\begin{lstlisting}[language=C++,numbers=left,xleftmargin=1.8em,label=KernelHookImplCode,caption={Implementation details of the data connection between a pull task and a kernel task.}]
struct PointerCaster {
  void* data {nullptr};
  template <typename T>  
  operator T* () {
    return (T*)data;
  }
};

template <typename T>
auto KernelTask::convert(T&& arg) {
  if constexpr(is_pull_task<T>) {
    return PointerCaster{arg.data()};
  }
  else {
    return forward<T>(arg);
  }
}
\end{lstlisting}

Each argument in the kernel function
must experience another conversion
(line 34 in Listing \ref{KernelTaskImplCode})
before launching the kernel.
The purpose of this conversion is to transform the pull task
to the type of the corresponding kernel argument,
and to possibly conduct any sanity checks at both compile time
and runtime.
Listing \ref{KernelHookImplCode}
highlights the core implementation of this conversion.
The function \texttt{convert} evaluates an argument at compile time
(line 9:17).
If the argument is a pull task, 
it returns a cast of the internal GPU data pointer to the target 
argument type (line 11:13).
Otherwise,
it forwards the argument in return
(line 15).
The auxiliary structure \texttt{PointerCaster} (line 1:7)
is designed to operate on plain old data (POD) pointers
in support for conventional GPU kernel programming syntaxes.
The same concept apply to custom data types
depending on a compiler's capability.

\subsubsection{Add a Dependency Link}

After tasks are created, the next step is to add dependency links.
A dependency link is a \textit{directed} edge between two tasks
to force one task to run before or after another.
Heteroflow defines two very intuitive methods, 
\texttt{precede} and \texttt{succeed},
to let users create task dependencies.
The two methods are symmetrical to each other.
A preceding link forces a task to run \textit{before} another and
a succeeding link forces a task to run \textit{after} another.
Heteroflow's task interface is uniform.
Users can insert dependencies between tasks of different types
as long as no cycles are formed.

\begin{figure}[h]
  \centering
  \centerline{\includegraphics[width=1.\columnwidth]{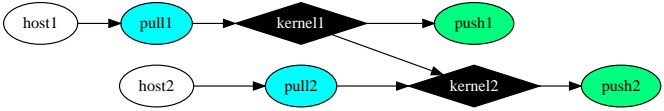}}
  \caption{A task graph of eight tasks and seven dependency constraints.}
  \label{fig::dependency}
\end{figure}

\begin{lstlisting}[language=C++,label=DependencyCode,caption={Creates dependency links to describe Figure \ref{fig::dependency}.}]
__global__ void k1(int* vec1);
__global__ void k2(int* vec1, int* vec2);

vector<int> vec1, vec2;

hf::Heteroflow hf;
auto host1 = hf.host([](){ vec1.resize(100, 0); });
auto host2 = hf.host([](){ vec2.resize(100, 1); });
auto pull1 = hf.pull(vec1);
auto pull2 = hf.pull(vec2);
auto push1 = hf.push(pull1, vec1);
auto push2 = hf.push(pull2, vec2);
auto kernel1 = hf.kernel(k1, pull1);
auto kernel2 = hf.kernel(k2, pull1, pull2);

host1.precede(pull1);
host2.precede(pull2);
pull1.precede(kernel1);
pull2.precede(kernel2);
kernel1.precede(push1, kernel2);
kernel2.precede(push2);
\end{lstlisting}

Listing \ref{DependencyCode} gives an example of using 
the method \texttt{precede} 
to describe the dependency graph in Figure \ref{fig::dependency}.
Users can precede an arbitrary number of tasks in one call.
The overall code to create dependency links in Heteroflow
is very \textit{simple}, \textit{concise}, and \textit{self-explanatory}.
An important takeaway here is that
task dependency is explicit in Heteroflow.
Our API never creates implicit dependency links 
even though they are obvious in certain graphs.
Such concern typically arises when creating a kernel task
that requires GPU data from other pull tasks.
In this scenario,
pull tasks must finish before the kernel task
and 
users are responsible for this dependency in their graphs.
Heteroflow delegates the dependency controls to users
so 
they can tailor graphs to their needs.
With careful graph construction and refinement,
applications can efficiently reuse data 
without adding redundant task dependencies.
For example, \texttt{kernel2} in Figure \ref{fig::dependency}
can access the GPU data of \texttt{pull1} 
as a result of transitive dependency 
(\texttt{pull1} precedes \texttt{kernel1} and
\texttt{kernel1} precedes \texttt{kernel2}).
Listing \ref{DependencyCode} implements this intent.

\subsubsection{Inspect a Task Dependency Graph}

Another powerful feature of Heteroflow on the user front
is the visualization of a task dependency graph
using the standard DOT format.
Users can find readily available tools such as
Python Graphviz and viz.js to draw a graph
without extra programming effort.
Graph visualization largely facilitates testing and debugging 
of Heteroflow applications.
Listing \ref{DebugCode} gives an example of dumping 
a Heteroflow graph to the standard output.

\begin{lstlisting}[language=C++,label=DebugCode,caption={Dumps a Heteroflow graph to the standard output.}]
hf.dump(cout);    
cout << hf.dump();
\end{lstlisting}

\subsection{Execute a Task Dependency Graph}

An \textit{executor} is the basic building block for executing 
a Heteroflow graph.
It manages a set of CPU threads and GPU devices
to \textit{schedule} in which list of tasks to execute.
When a task is ready,
the runtime submits the task to an \textit{execution context}
which can occur in either a physical CPU core or a GPU device.
In Heteroflow, a task is indeed a callable.
When users create a task,
Heteroflow marshals all required parameters along with 
unique placeholders for runtime arguments 
to form a closure that can be run by any CPU thread.
Execution of a GPU task will be placed under a GPU context.
The scheduler manages all such details 
to ensure consistent results across multiple GPUs.
Listing \ref{ExecutorCode}
creates an executor of eight CPU threads and four GPUs
and uses it to execute a graph one times, 100 times,
and multiple times until a stopping criteria is met.
Users can adjust the number based on hardware capability
to easily scale their graphs across different CPU-GPU configurations.
All the run methods in the executor class 
are \textit{non-blocking}.
Issuing a run on a graph returns immediately with 
a C++ \textit{future} object.
Users can use it to inspect the execution status of the graph
or chain up a continuation for asynchronous controls.
The executor class also provides a method \texttt{wait\_for\_all} that blocks 
until all running graphs associated with the caller executor 
finish.
Heteroflow's executor interface is \textit{thread-safe}.
Touching an executor from multiple threads is valid.
Users can take advantage of this property to 
explore higher-level parallelism 
without concerning about race in execution.
  
\begin{lstlisting}[language=C++,label=ExecutorCode,caption={Creates an executor to run a Heteroflow graph.}]
hf::Executor executor(8, 4); // 8 CPU threads 4 GPUs
hf::Heteroflow graph;
auto future1 = executor.run(graph);
auto future2 = executor.run_n(graph, 100);
auto future3 = executor.run_until(graph, [&] () {
  return custom_stopping_criteria();
});
executor.wait_for_all();
\end{lstlisting}

\subsection{Scheduling Algorithm}

Another major contribution of Heteroflow is the design of a
\textit{scheduler} on top of our heterogeneous tasking interface.
Scheduler is an integral part of the executor 
for mapping task graphs onto available CPU cores and GPUs.
When an executor is created with $N$ CPU threads and $M$ GPUs,
we spawn $N$ CPU threads, namely \textit{workers}, to execute tasks.
Unlike existing works~\cite{StarPU, XKAAPI++}, 
we do not dedicate a worker to manage a target GPU,
since all tasks are uniformly represented in Heteroflow 
using polymorphic functional objects
(see Listings \ref{PullTaskImplCode}, 
\ref{PushTaskImplCode}, and \ref{KernelTaskImplCode}).
This largely facilitates the design of our scheduler 
in providing efficient resource utilization
and flexible runtime optimizations,
for instance, GPU memory allocators, asynchronous CUDA streams,
and task fusing.

Our scheduler design is motivated by~\cite{Huang_19_01}.
When a graph is submitted to an executor,
a special data structure called \textit{topology} is created 
to marshal execution parameters and runtime metadata.
Each heteroflow object has a list of topologies to
track individual execution status.
The executor also maintains a topology counter 
to signal callers on completion.
The communication is based on a shared state managed by a pair of C++
\textit{promise} and \textit{future} objects.
The first step in scheduling is \textit{device placement},
mapping each GPU task to a particular GPU device.
An advantage of our programming model is implicit data dependencies
between a kernel and its pull tasks (see line 3 in Listing \ref{KernelTaskImplCode}),
through which the scheduler can utilize to place them under the right device.
Based on this property,
we develop a simple and efficient device placement algorithm
using \textit{union-find} and \textit{bin packing} 
as shown in Algorithm \ref{alg::DevicePlacement}.
The key idea is to group each kernel with its source pull tasks
(line 1:7)
and then pack each unique group to a GPU bin with an optimized cost
(line 8:14).
By default, we minimize the load per GPU bins for maximal concurrency
but can expose this strategy
to a pluggable interface for custom cost metrics.

\begin{algorithm}[h]
  \SetAlgoLined
  \SetKw{false}{false}
  \SetKw{true}{true}
  \SetKw{nullopt}{nullopt}
  \SetKw{auto}{auto}
  \SetKw{or}{or}
  \ForEach{t $\in$ tasks} {
    \If{t.type() $==$ KERNEL}{
      \ForEach{p $\in$ t.source\_pull\_tasks()} {
        set\_union($t$, $p$);
      }
    }
  }
  \ForEach{t $\in$ tasks} {
    \If{x $\leftarrow$ t.type(); x == KERNEL \or x == PULL} {
      \If{$r \leftarrow$ set\_find($t$); is\_set\_root(r) } {
        set\_bin\_packing\_with\_balanced\_load($t$);
      }
    }
  }
\caption{DevicePlacement}
\label{alg::DevicePlacement}
\end{algorithm}

After device placement,
the scheduler enters a \textit{work-stealing} loop where
each worker thread iteratively drains out tasks from its local queue
and transitions to a \textit{thief} to steal a task 
from a randomly selected peer called \textit{victim}.
The process stops when an executor is destroyed.
We employ work-stealing because it has been extensively studied and used in
many parallel processing systems 
for dynamic load-balancing and irregular computations~\cite{Lima_15_01, Lin_20_01}.
When a worker thread executes a task,
it applies a \textit{visitor} pattern that invokes a separate method 
for each task type.
Running a host task is trivial, 
but calling a GPU task must be scoped under the right execution context.
Heteroflow provides a \textit{resource acquisition is initialization} (RAII)-style
mechanism on top of CUDA device API to scope the task execution
under its assigned GPU device. 
Listing \ref{ScopedPullTask} gives the implementation details of
invoking a pull task from an executor.
All GPU tasks are synchronized through CUDA events
(line 4 and line 6).

\begin{lstlisting}[language=C++,numbers=left,xleftmargin=1.8em,label=ScopedPullTask,caption={Implementation details of invoking a pull task.}]
void Executor::invoke(unsigned me, Pull& h) {
  auto [d, s, e] = get_device_stream_event(me, h);
  ScopedDeviceContext ctx(d);
  cudaEventRecord(e, s);
  h.work(get_device_allocator(d), s);
  cudaStreamWaitEvent(s, e, 0);
}
\end{lstlisting}

While detailing the scheduler design is out of the scope of this paper,
there are a few notable items.
First, each worker keeps a \textit{per-thread} CUDA stream
to enable concurrent GPU memory and kernel operations.
Second, our executor keeps a \textit{memory pool}
for each GPU device to reduce the scheduling overhead 
of frequent allocations by pull tasks.
We implement the famous Buddy allocator algorithm~\cite{Buddy}.
Third, our work-stealing loop adopts an adaptive strategy 
to balance working and sleeping threads 
on top of available task parallelism.
The key idea is to ensure one thief exists as long as 
an active worker is running a task.
At the time of this writing, 
our scheduler design might not be perfect, 
but it provides a proof of concept for our programming model
and fosters future research opportunities for 
new algorithms.
%

\section{Experimental Results}

We evaluated the performance of Heteroflow on two real VLSI CAD applications,
timing analysis and standard cell placement.
Each application represents a unique computation pattern.
All experiments ran on a Ubuntu Linux 5.0.0-21-generic x86 64-bit 
machine with 40 Intel Xeon Gold 6138 CPU cores at 2.00 GHz,
4 GeForce RTX 2080 GPUs, and 256 GB RAM.
The timing analysis program is compiled by g++8.2 
and nvcc CUDA 10.1 with C++14 standards \texttt{-std=c++14} 
and optimization flags \texttt{-O2}.
The placement program is compiled under the same environment.
Both programs are derived from our open-source projects,
OpenTimer~\cite{Huang_15_01, Huang_21_01, Huang_21_02} and DREAMPlace~\cite{DREAMPlace},
that consist of complex domain-specific algorithms 
with more than 10K lines of code
over years of development.
%

%
\subsection{VLSI Timing Analysis}

\begin{figure}[h]
  \centering
  \centerline{\includegraphics[width=1.\columnwidth]{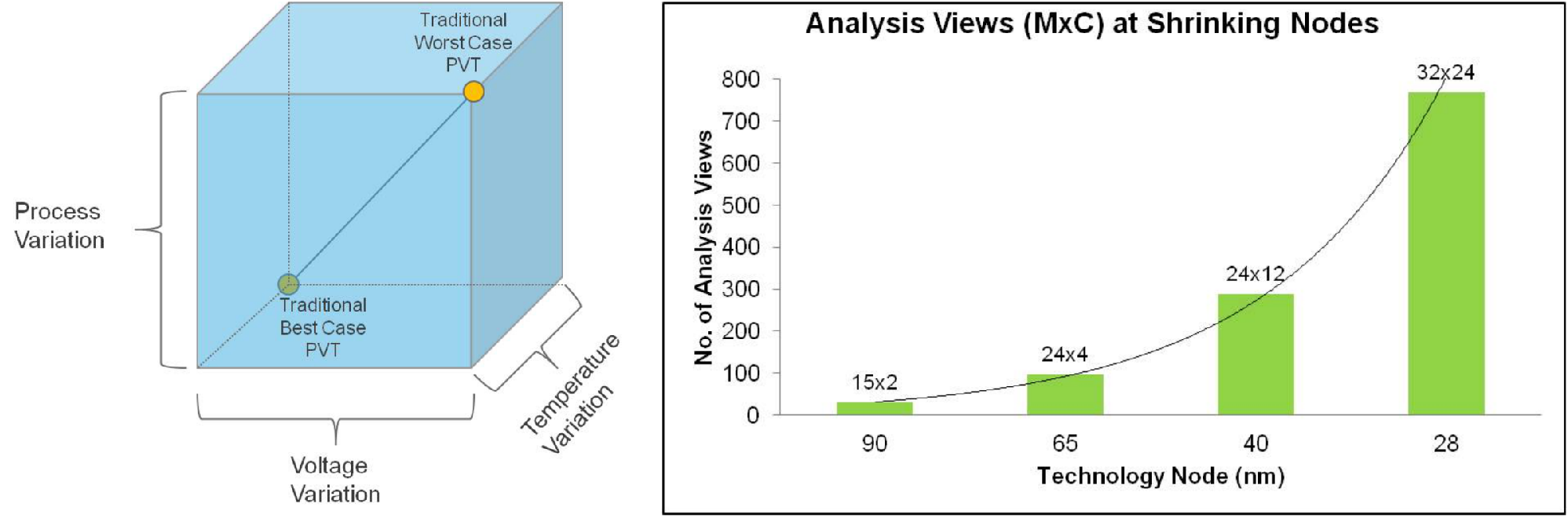}}
  \caption{The required analysis views in terms of corners and modes increase exponentially as the technology node advances.}
  \label{fig::timing-views}
\end{figure}

We applied Heteroflow to solve a VLSI timing analysis problem.
Timing analysis is a very important component 
in the overall design flow (see Figure \ref{fig::cad}).
It verifies the expected timing behaviors of a digital circuit
to ensure correct functionalities after tape-out.
Among various timing analysis problems,
one subject is to find the correlation between different
\textit{timing views}.
Each each view represents a unique
combination of a process variation corner (e.g., temperature, voltage)
and an analysis mode (e.g., testing, functional).
Figure \ref{fig::timing-views} shows the number of required analysis views
increases exponentially as the technology node advances~\cite{Huang_15_01, Huang_21_01}.
Timing correlation is not only important 
for reasoning the behavior of a timer
but also useful for building regression models to 
reduce required analysis iterations.

\begin{figure}[h]
  \centering
  \centerline{\includegraphics[width=1.\columnwidth]{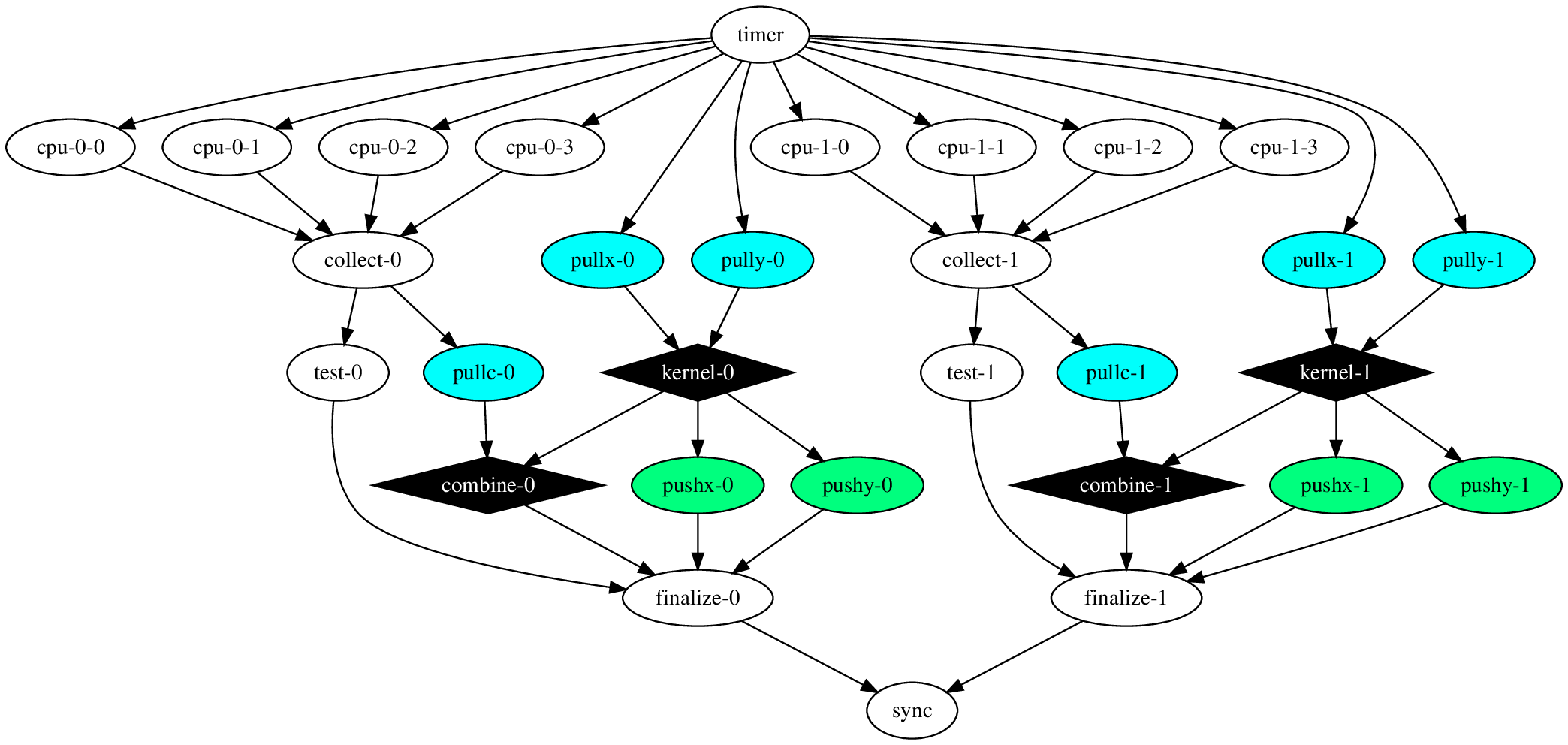}}
  \caption{A partial task graph of VLSI timing analysis for finding correlation between two views. Each view implements a hybrid CPU-GPU correlation algorithm.}
  \label{fig::exp-timing-task-graph}
\end{figure}

In reality, there are many ways to conduct timing analysis and correlation.
In this experiment, we consider a representative three-step flow:
a timer generates analysis datasets from a circuit design
across multiple views;
a hybrid CPU-GPU algorithm extracts timing statistics 
and generates regression models for each dataset;
a synchronization step combines all assessed quantities 
to a concrete report.
Figure \ref{fig::exp-timing-task-graph} illustrates
a fractional task graph of two views.
We use the open-source tool, OpenTimer, 
to generate 1024 different timing reports 
for a large circuit, \textit{netcard}, of 1.5M gates~\cite{Huang_15_01, Huang_21_01}.
The correlation layer implements a CPU-based algorithm to extract 
graph information (critical paths~\cite{HPEC_21_01, Zhou_22_01}, CPPR~\cite{Huang_14_01, Huang_16_01, HeteroCPPR}) and 
a GPU-based algorithm to perform logistic regression
with gradient descent.
Part of CPU and GPU tasks are dependent to each other.
For demonstration purpose,
we pre-generate the analysis data
and control the sample size such that each analysis view
takes approximately the same runtime.

\input{Fig/exp-timing/exp-timing.tex}

Figure \ref{fig::exp-timing-strong-scaling} shows the overall runtime performance 
at different CPU-GPU numbers and problem sizes.
In general, we observe a descent scaling when increasing the number
of cores and GPUs.
The task graph requires 99 minutes to finish at
the lowest hardware concurrency of 1 core and 1 GPU.
Using all 40 cores and 4 GPUs is able to speed up the runtime 
by $7.7\times$ finishing in 13 minutes.
On the slice of 4 GPUs,
the runtimes are 51, 23, 18, 15, 14, and 13 minutes
for 1, 8, 16, 24, 32, and 40 cores, respectively.
The GPU counterparts at 40 cores 
are 36, 21, 15, and 13 minutes 
for 1, 2, 3, and 4 GPUs, respectively.   
The lower side of Figure \ref{fig::exp-timing-strong-scaling} shows
the runtime versus the problem size in terms of
six different timing views, 32, 64, 128, 256, 512, and 1024.
At any point,
increasing the number of CPUs or GPUs can all reduce the runtime.
For this particular workload,
speed-up from multiple GPUs is more remarkable than CPUs.

\subsection{VLSI Placement}

\begin{figure}[tb]
    \centering
    \subfloat[]{\includegraphics[width=.4\columnwidth]{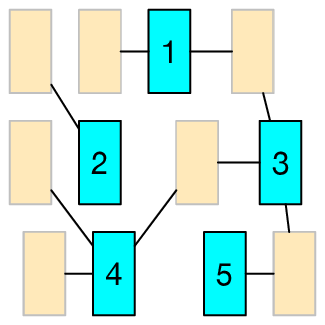}}\hspace{.1in}
    \subfloat[]{\includegraphics[width=.4\columnwidth]{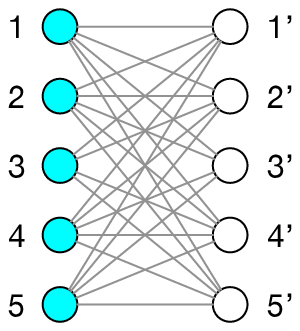}}\\
    \subfloat[]{\includegraphics[width=.9\columnwidth]{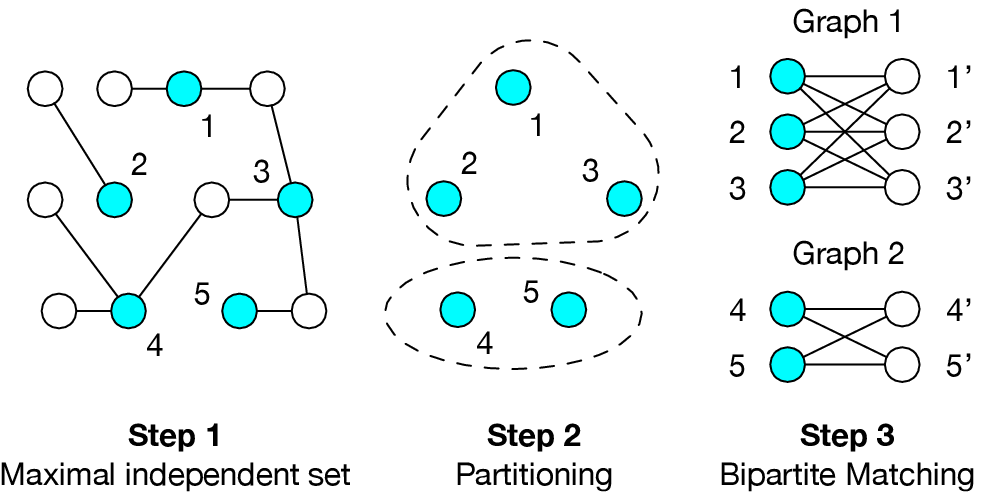}}
    \caption{
        A matching-based detailed placement algorithm.
        (a) A placement example of cells and interconnects. Independent cells are marked in cyan.
        (b) A weighted bipartite matching formulation to find the best permutation of cell locations. 
        (c) A practical three-step iterative implementation of the algorithm. 
    }
    \label{fig::place}
\end{figure}

We applied Heteroflow to solve a VLSI placement problem,
a fundamental step in the physical design stage 
(see Figure \ref{fig::cad}).
The goal is to determine the physical locations of 
cells (logic gates) in a fixed layout region
with minimal interconnect wirelength.
Modern placement typically incorporates hundreds of millions of cells
and takes several hours to finish.
To reduce the long runtime, 
recent work started investigating new algorithms 
using the power of heterogeneous computing~\cite{DREAMPlace}.
Among various placement techniques, 
\textit{detailed placement} is an important step
to refine a legalized placement solution 
for minimal wirelength.
Mainstream detailed placement algorithms are
combinatorial and iterative.
A widely-used matching-based algorithm is shown in Figure \ref{fig::place}.
The key idea is to extract a maximal independent set
(marked in cyan) from a cell set 
and model the wirelength minimization problem
on these non-overlapped cells 
into a weighted bipartite matching graph. 
The entire process is very time-consuming
especially for large designs with millions of cells.
A practical implementation iterates the following three steps:
a parallel maximal independent set finding step using 
Blelloch's Algorithm~\cite{blelloch2012greedy};
a sequential partitioning step to cluster adjacent cells;
a parallel bipartite matching step to find the best permutation of cell locations. 
Figure \ref{fig::place}(c) illustrates the process.

\begin{figure}[tb]
    \centering
    \includegraphics[width=1.\columnwidth]{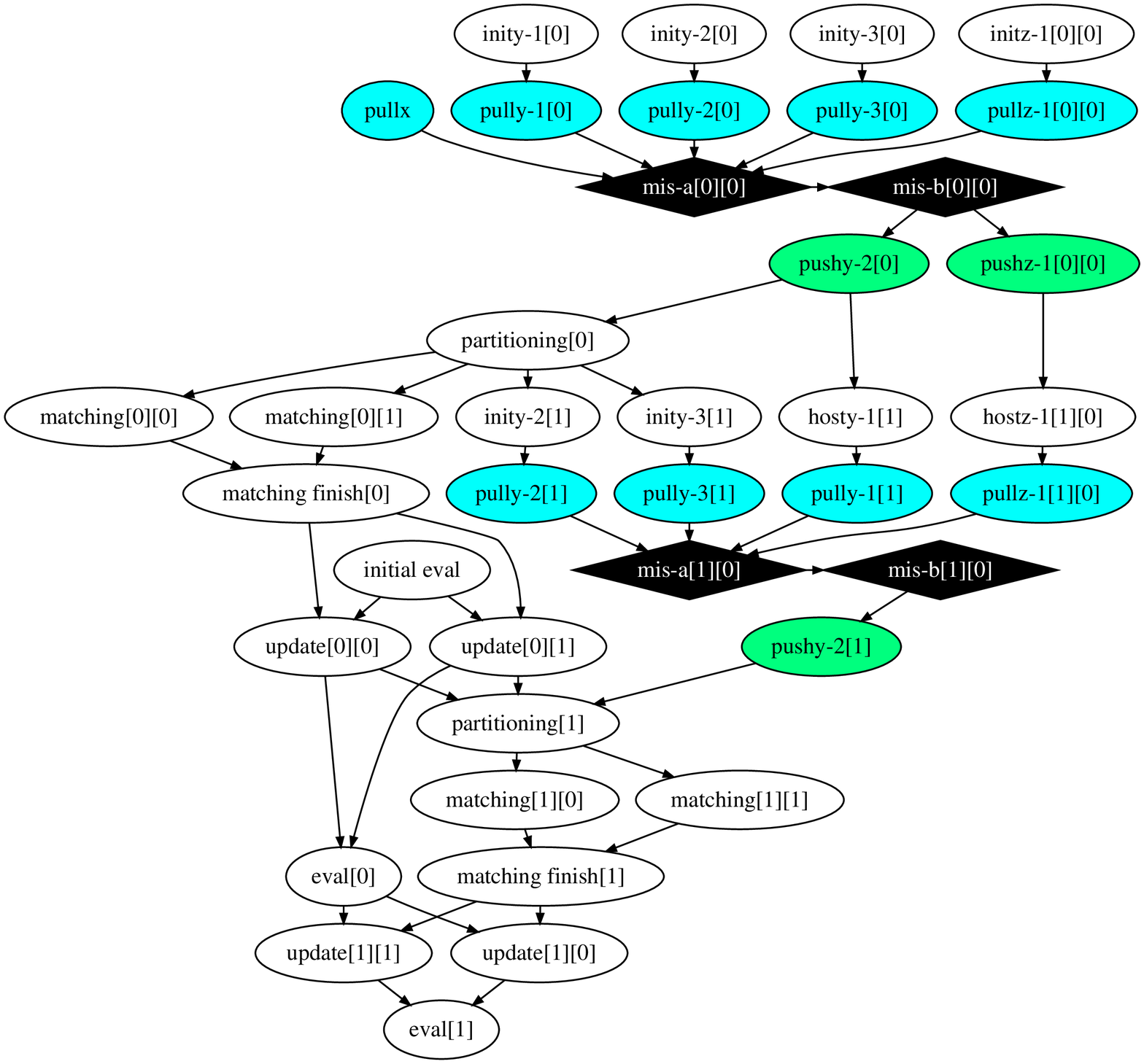}
    \caption{
        A partial task graph for the detailed placement algorithm
        of two iterations. The number in a square bracket indicates the iteration number.
    }
    \label{fig::place-dependency}
\end{figure}

In the experiment, we implemented a hybrid CPU-GPU 
detailed placement algorithm introduced by DREAMPlace~\cite{DREAMPlace}.
Among these three steps,
finding the maximal independent set takes the most runtime.
DREAMPlace developed a new acceleration algorithm that offloaded this step
to GPU, and showed $40\times$ speed-up 
over a CPU baseline using 20 cores~\cite{DREAMPlace}.
The other two steps have graph-oriented computation patterns 
and are implemented on CPUs.
Figure \ref{fig::place-dependency} shows a partial task graph 
for the algorithm in two iterations.
The algorithm normally converges in 10-50 iterations.
To enable task overlaps between iterations,
we flatten the task graph for a given iteration number.
The task graph in Figure \ref{fig::place-dependency} 
highlights the complexity of the algorithm
and dependent CPU-GPU tasks.

\input{Fig/exp-placement/exp-placement.tex}

Figure \ref{fig::exp-placement-strong-scaling} shows the runtime performance
at different CPU-GPU numbers and iterations,
for placing a large circuit, \textit{bigblue4},
of 2.2M cells and 2.2M nets. 
It is observed that increasing the number of CPU cores reduces the runtime.
For instance,
under 1 GPU it takes 58.41s and 14.02s using 1 core and 40 cores,
respectively.
Maximum concurrency saturates at approximately 20 cores.
In this particular workload,
performance with 1 GPU is good enough.
The runtime does not benefit too much from adding more GPUs.
We can clearly see this property on the upper-right plot.
Under 40 cores, it takes 14.02s and 13.61s for 1 GPU and 4 GPUs, respectively.
In fact, this property is generally true for most optimization algorithms
in VLSI CAD,
as they are often irregular and dependent~\cite{Lu_18_01}.
In terms of different problem sizes which is
measured by the iteration count used to construct the task graph,
increasing the number of CPU cores can reduce the runtime
in most scenarios.
For example, the task graph of 5 iterations under 4 GPUs
finishes in 6.35s and 1.44s using 1 core and 40 cores, respectively.
Due to the nature of the algorithm, 
such trend is not observed on the GPU side.

\section{Related Work}

\textit{\textbf{Heterogeneous programming models}} have been extensively
developed in scientific communities 
and enabled vast success in various problem domains~\cite{Mittal_15_01}.
CUDA, OpenCL, OpenGL, C++ AMP, and Brook+
are popular GPU programming frameworks that provide
a rich set of low-level APIs for 
explicit GPU managements~\cite{CUDA, OpenCL, OpenGL, Brook}.
These libraries are designed particularly for power users
to implement various optimization strategies specific to 
a GPU architecture.
Directive-based models such as 
hiCUDA, Ompss, OpenMPC, and OpenACC
provide high-level abstraction
on GPU programming by
augmenting program information,
for instance, guidance on loop mapping onto GPU and
data sharing rules,
to designated compilers~\cite{hiCUDA, Ompss, OpenMPC, OpenACC}.
These models are good at loop-based parallelism
but cannot handle well 
irregular task parallelism~\cite{Lee_12_01}.
Functional-level approaches such as
StarPU, SYCL, HPX, PaRSEC, QUARK, XKAAPI++, Unicorn, and Taskflow
are capable of concurrent CPU-GPU tasking~\cite{StarPU, SYCL, HPX, PaRSEC, XKAAPI++, QUARK, Beri_16_01, Huang_22_01, Lin_19_01, Lin_19_02}.
The offered graph description languages 
can be complex or expressive, depending on the targeted applications.
Other data structure-driven libraries 
such as Thrust, VexCL, and Boost.Compute
provide C++ STL-style interfaces to program 
batch CPU-GPU workloads~\cite{Thrust, VexCL, BoostCompute}.
For concurrent CPU-GPU tasking,
users are responsible for scheduling
and concurrency controls that are known difficult 
to program correctly.

\textit{\textbf{CPU-GPU co-scheduling}} 
is a pivotal component of all heterogeneous programming systems.
The parallel computing community has a number of
algorithms including static mapping~\cite{Scogland_12_01},
dynamic work-stealing~\cite{Lima_15_01, Lin_20_01},
asymptotic profiling~\cite{Wang_14_01}, and
other system-defined strategies~\cite{Ompss, StarPU, HPX, Beri_16_01}.
Vendor-specific features such as CUDA Graph~\cite{CUDA, cudaFlow}
and SYCL~\cite{SYCL} offer asynchronous graph scheduling
for task parallelism but implementation details are unknown.
On the other hand,
automatic GPU placement has been studied in 
machine learning community~\cite{Mirhoseini_18_01, Gao_18_01}.
The goal is to place operations in a deep neural network onto GPU devices 
in an optimal way, such that
the training process can complete within the shortest amount of time.
However, these algorithms are problem-specific 
and require a unified tensor data structure 
for performance modeling.
%


\section{Conclusion}

In this paper, we have introduced Heteroflow,
a new modern C++ tasking library to help developers quickly write
CPU-GPU parallel programs and implement efficient
heterogeneous decomposition algorithms.
We have evaluated Heteroflow on two real design automation problems
and shown performance scalability across different CPU-GPU numbers
and problem sizes.
At a particular VLSI timing analysis example,
Heteroflow can reduce a baseline runtime from 99 minutes 
to 13 minutes ($7.7 \times$ speed-up) on a machine of 40 CPU cores and 4 GPUs.
Future work will focus on distributing our scheduler based on~\cite{DtCraft}
and incorporating a broader range of workloads, including machine learning~\cite{SNIG_HPEC_20, SNIG_TPDS_22}
and engineering simulation~\cite{GPUGBA, GPUPBA, Huang_21_03}.

\bibliographystyle{IEEEtran}
\bibliography{IEEEabrv,ms}

\end{document}

%% file: Fig/exp-timing/exp-timing.tex
\begin{figure}[h]
  \centering

  \pgfplotsset{
    title style={font=\Large},
    label style={font=\large},
  }
  \begin{tikzpicture}[scale=0.51]
    \begin{axis}[
      title=Runtimes across CPUs,
      xlabel=Number of CPU Cores,
      ylabel=Runtime (minutes),
      xtick={1, 8, 16, 24, 32, 40},
      legend pos=north east,
    ]
    \addplot+ table[x=cores,y=four-gpus,col sep=space]{Fig/exp-timing/cpu-scalability.csv};
    \addplot+ table[x=cores,y=three-gpus,col sep=space]{Fig/exp-timing/cpu-scalability.csv};
    \addplot+ table[x=cores,y=two-gpus,col sep=space]{Fig/exp-timing/cpu-scalability.csv};
    \addplot+ table[x=cores,y=one-gpu,col sep=space]{Fig/exp-timing/cpu-scalability.csv};
    \legend{4 GPUs, 3 GPUs, 2 GPUs, 1 GPU}
    \end{axis}
  \end{tikzpicture}
  \begin{tikzpicture}[scale=0.51]
    \begin{axis}[
      title=Runtimes across GPUs,
      xlabel=Number of GPUs,
      ylabel=Runtime (minutes),
      xtick={1, 2, 3, 4},
      legend pos=north east,
    ]
    \addplot+ table[x=gpus,y=40,col sep=space]{Fig/exp-timing/gpu-scalability.csv};
    \addplot+ table[x=gpus,y=32,col sep=space]{Fig/exp-timing/gpu-scalability.csv};
    \addplot+ table[x=gpus,y=24,col sep=space]{Fig/exp-timing/gpu-scalability.csv};
    \addplot+ table[x=gpus,y=16,col sep=space]{Fig/exp-timing/gpu-scalability.csv};
    \addplot+ table[x=gpus,y=8,col sep=space]{Fig/exp-timing/gpu-scalability.csv};
    \legend{40 CPU cores, 32 CPU cores, 24 CPU cores, 16 CPU cores, 8 CPU cores}
    \end{axis}
  \end{tikzpicture}
  \begin{tikzpicture}[scale=0.50]
    \begin{axis}[
      title=Runtime vs Problem Size (40 cores),
      xlabel=Problem Size (\# analysis views),
      ylabel=Runtime (minutes),
      legend pos=north west,
    ]
    \addplot+ table[x=N,y=four-gpus,col sep=space]{Fig/exp-timing/weak-scalability-gpu.csv};
    \addplot+ table[x=N,y=three-gpus,col sep=space]{Fig/exp-timing/weak-scalability-gpu.csv};
    \addplot+ table[x=N,y=two-gpus,col sep=space]{Fig/exp-timing/weak-scalability-gpu.csv};
    \addplot+ table[x=N,y=one-gpu,col sep=space]{Fig/exp-timing/weak-scalability-gpu.csv};
    \legend{4 GPUs, 3 GPUs, 2 GPUs, 1 GPU}
    \end{axis}
  \end{tikzpicture}
  \begin{tikzpicture}[scale=0.50]
    \begin{axis}[
      title=Runtime vs Problem Size (4 GPUs),
      xlabel=Problem Size (\# analysis views),
      ylabel=Runtime (minutes),
      legend pos=north west,
    ]
    \addplot+ table[x=N,y=40,col sep=space]{Fig/exp-timing/weak-scalability-cpu.csv};
    \addplot+ table[x=N,y=32,col sep=space]{Fig/exp-timing/weak-scalability-cpu.csv};
    \addplot+ table[x=N,y=24,col sep=space]{Fig/exp-timing/weak-scalability-cpu.csv};
    \addplot+ table[x=N,y=16,col sep=space]{Fig/exp-timing/weak-scalability-cpu.csv};
    \addplot+ table[x=N,y=8,col sep=space]{Fig/exp-timing/weak-scalability-cpu.csv};
    \legend{40 Cores, 32 Cores, 24 Cores, 16 Cores, 8 Cores}
    \end{axis}
  \end{tikzpicture}
  \caption{Runtimes at different CPU-GPU numbers and problem sizes for analyzing the circuit netcard (1.5M gates and 1.5M nets).}
  \label{fig::exp-timing-strong-scaling}
\end{figure}

%% file: Fig/exp-placement/exp-placement.tex
\begin{figure}[h]
  \centering

  \pgfplotsset{
    title style={font=\Large},
    label style={font=\large},
  }
  \begin{tikzpicture}[scale=0.51]
    \begin{axis}[
      title=Runtimes across CPUs,
      xlabel=Number of CPU Cores,
      ylabel=Runtime (seconds),
      xtick={1, 8, 16, 24, 32, 40},
      legend pos=north east,
    ]
    \addplot+ table[x=cores,y=four-gpus,col sep=space]{Fig/exp-placement/cpu-scalability.csv};
    \addplot+ table[x=cores,y=three-gpus,col sep=space]{Fig/exp-placement/cpu-scalability.csv};
    \addplot+ table[x=cores,y=two-gpus,col sep=space]{Fig/exp-placement/cpu-scalability.csv};
    \addplot+ table[x=cores,y=one-gpu,col sep=space]{Fig/exp-placement/cpu-scalability.csv};
    \legend{4 GPUs, 3 GPUs, 2 GPUs, 1 GPU}
    \end{axis}
  \end{tikzpicture}
  \begin{tikzpicture}[scale=0.51]
    \begin{axis}[
      title=Runtimes across GPUs,
      xlabel=Number of GPUs,
      ylabel=Runtime (seconds),
      xtick={1, 2, 3, 4},
      legend pos=north east,
      legend style={at={(0.95,0.80)}}
    ]
    \addplot+ table[x=gpus,y=40,col sep=space]{Fig/exp-placement/gpu-scalability.csv};
    \addplot+ table[x=gpus,y=32,col sep=space]{Fig/exp-placement/gpu-scalability.csv};
    \addplot+ table[x=gpus,y=24,col sep=space]{Fig/exp-placement/gpu-scalability.csv};
    \addplot+ table[x=gpus,y=16,col sep=space]{Fig/exp-placement/gpu-scalability.csv};
    \addplot+ table[x=gpus,y=8,col sep=space]{Fig/exp-placement/gpu-scalability.csv};
    \addplot+ table[x=gpus,y=1,col sep=space]{Fig/exp-placement/gpu-scalability.csv};
    \legend{40 CPU cores, 32 CPU cores, 24 CPU cores, 16 CPU cores, 8 CPU cores, 1 CPU core}
    \end{axis}
  \end{tikzpicture}
  \begin{tikzpicture}[scale=0.50]
    \begin{axis}[
      title=Runtime vs Problem Size (40 cores),
      xlabel=Problem Size (\# iterations),
      ylabel=Runtime (seconds),
      legend pos=north west,
    ]
    \addplot+ table[x=N,y=four-gpus,col sep=space]{Fig/exp-placement/weak-scalability-gpu.csv};
    \addplot+ table[x=N,y=three-gpus,col sep=space]{Fig/exp-placement/weak-scalability-gpu.csv};
    \addplot+ table[x=N,y=two-gpus,col sep=space]{Fig/exp-placement/weak-scalability-gpu.csv};
    \addplot+ table[x=N,y=one-gpu,col sep=space]{Fig/exp-placement/weak-scalability-gpu.csv};
    \legend{4 GPUs, 3 GPUs, 2 GPUs, 1 GPU}
    \end{axis}
  \end{tikzpicture}
  \begin{tikzpicture}[scale=0.50]
    \begin{axis}[
      title=Runtime vs Problem Size (4 GPUs),
      xlabel=Problem Size (\# iterations),
      ylabel=Runtime (seconds),
      legend pos=north west,
    ]
    \addplot+ table[x=N,y=40,col sep=space]{Fig/exp-placement/weak-scalability-cpu.csv};
    \addplot+ table[x=N,y=32,col sep=space]{Fig/exp-placement/weak-scalability-cpu.csv};
    \addplot+ table[x=N,y=24,col sep=space]{Fig/exp-placement/weak-scalability-cpu.csv};
    \addplot+ table[x=N,y=16,col sep=space]{Fig/exp-placement/weak-scalability-cpu.csv};
    \addplot+ table[x=N,y=8,col sep=space]{Fig/exp-placement/weak-scalability-cpu.csv};
    \addplot+ table[x=N,y=1,col sep=space]{Fig/exp-placement/weak-scalability-cpu.csv};
    \legend{40 Cores, 32 Cores, 24 Cores, 16 Cores, 8 Cores, 1 Core}
    \end{axis}
  \end{tikzpicture}
  \caption{Runtimes at different CPU-GPU numbers and problem sizes for placing the circuit bigblue4 (2.2M cells and 2.2M nets).}
  \label{fig::exp-placement-strong-scaling}
\end{figure}